\documentclass[sigconf]{acmart}

\copyrightyear{2026}
\acmYear{2026}
\setcopyright{cc}
\setcctype{by}
\acmConference[ICSE '26]{2026 IEEE/ACM 48th International Conference on Software Engineering}{April 12--18, 2026}{Rio de Janeiro, Brazil}
\acmBooktitle{2026 IEEE/ACM 48th International Conference on Software Engineering (ICSE '26), April 12--18, 2026, Rio de Janeiro, Brazil}
\acmDOI{10.1145/3744916.3773192}
\acmISBN{979-8-4007-2025-3/26/04}

\usepackage{amsmath, amsfonts}
\usepackage{textcomp}
\usepackage{enumitem}

\usepackage{booktabs}
\usepackage{balance}
\usepackage{algorithm}
\usepackage{algorithmic}
\usepackage{xspace}
\usepackage{multirow}
\usepackage{soul}

\usepackage{pifont}

\newenvironment{mybox}{
    \begin{center}
    \begin{minipage}{0.48\textwidth}
    \hrule height 0.8pt
    \vspace{4pt}
    \normalsize
    \begin{list}{}{
        \leftmargin=6pt
        \rightmargin=6pt
    }
    \item[]
}{
    \end{list}
    \vspace{4pt}
    \hrule height 0.8pt
    \end{minipage}
    \end{center}
}

\usepackage[english]{babel}

\theoremstyle{definition}

\def\BibTeX{{\rm B\kern-.05em{\sc i\kern-.025em b}\kern-.08em
    T\kern-.1667em\lower.7ex\hbox{E}\kern-.125emX}}

\begin{document}

\title{What Makes Code Generation Ethically Sourced?}


\settopmatter{authorsperrow=4}
\author{Zhuolin Xu}
\affiliation{%
    \institution{Concordia University}
    \city{Montreal}
    \country{Canada}
}
\email{zhuolin.xu@mail.concordia.ca}
\author{Chenglin Li}
\affiliation{%
    \institution{Concordia University}
    \city{Montreal}
    \country{Canada}
}
\email{chenglin.li@mail.concordia.ca}
\author{Qiushi Li}
\affiliation{%
    \institution{Concordia University}
    \city{Montreal}
    \country{Canada}
}
\email{qiushi.li@mail.concordia.ca}
\author{Shin Hwei Tan}
\authornote{Corresponding Author}
\affiliation{%
    \institution{Concordia University}
    \city{Montreal}
    \country{Canada}
}
\email{shinhwei.tan@concordia.ca}




\newcommand{\tooln}{\textsc{ES-CodeGen}\xspace} 
\newcommand{\numdim}{13\xspace} 
\newcommand{\numdomain}{10\xspace} 
\newcommand{\numreply}{34\xspace}
\newcommand{\manualread}{803\xspace}
\newcommand{\relevantpaper}{71\xspace}
\newcommand{\numreplyvalid}{32\xspace}
\newcommand{\numpart}{20\xspace} 
\newcommand{\numrefactor}{32\xspace}
\newcommand{\avgaccuracy}{97.55\%\xspace}
\newcommand{\avggn}{65.93\%\xspace}

\newcommand{\chmark}{\ding{51}}%
\newcommand{\xhmark}{\ding{55}}%
\newcommand{\numcomptool}{five\xspace} 
\newcommand{\numcompopen}{four\xspace} 
\newcommand{\datasetsize}{100} 
\newcommand{\cptemp}{Diversity-enhanced Prompt Synthesis\xspace}
\newcommand{\ptemp}{diversity-enhanced prompt synthesis\xspace}
\newcommand{\bench}{coverage-guided harmful content dataset\xspace}
\newcommand{\cbench}{Coverage-guided Harmful Content Dataset\xspace}
\newcommand{\jdeo}{\textsc{JDeodorant}\xspace}

\newcommand{\tufano}{\textsc{TufanoNMT}\xspace}
\newcommand{\autotransform}{\textsc{AutoTransform}\xspace}
\newcommand{\Llama}{\textsc{Code Llama}\xspace}
\newcommand{\eclipse}{\textsc{Eclipse}\xspace}
\newcommand{\idea}{\textsc{IntelliJ IDEA}\xspace}
\newcommand{\jdt}{\textsc{JDT}\xspace}
\newcommand{\refactoringtypeinpreviousdataset}{13}

\definecolor{modifyhaibo}{rgb}{0.0, 0.5, 1.0}

\newcommand{\totnumprojbench}{25022\xspace} 
\newcommand{\totnumreport}{eight\xspace} 
\newcommand{\totnumfixed}{seven\xspace}
\newcommand{\totnumconfirmed}{one\xspace} 
\newcommand{\totnumaccept}{four\xspace}

\newcommand{\numbench}{329\xspace} 
\newcommand{\numprojbench}{307\xspace} 

\newcommand{\truepositive}{87\xspace} 

\newcommand{\exactmatch}{30\xspace} 

\newcommand{\studycohenkappa}{0.81\xspace} 

\newcommand{\nopattern}{eight} 

\newcommand{\Plus}{\mathord{\begin{tikzpicture}[baseline=0ex, line width=0.5, scale=0.06]\draw (1,0) -- (1,2);
\draw (0,1) -- (2,1);
\end{tikzpicture}}}
\newcommand{\Minus}{\mathord{\begin{tikzpicture}[baseline=0ex, line width=0.5, scale=0.06]
\draw (0,1) -- (2,1);
\end{tikzpicture}}}
\def\HiLir{\leavevmode\rlap{\hbox to \hsize{\color{red!50}\leaders\hrule height .8\baselineskip depth .5ex\hfill}}}
\def\HiLi{\leavevmode\rlap{\hbox to \hsize{\color{blue!50}\leaders\hrule height .8\baselineskip depth .5ex\hfill}}}

\newcommand\FIXME[1]{\textcolor{red}{#1}}
\newcommand{\delcatstudy}{16.49}
\newcommand{\modifyshin}[1]{{\color{red}{#1}}}
\newcommand{\simpli}{developer-induced program simplification}
\newcommand{\csimpli}{Developer-induced Program Simplification}
\newcommand{\cpsimpli}{Developer-induced program simplification}
\definecolor{azure(colorwheel)}{rgb}{0.0, 0.5, 1.0}

\newcommand{\useapip}{Replace with equivalent API}
\newcommand{\useapi}{``\useapip''\xspace}

\newcommand{\refactoringminer}{RefactoringMiner}

\newcommand{\evalproj}{eights}

\newcommand{\evalbugs}{36}
\newcommand{\origge}{54}
\newcommand{\geevalbugs}{42}
\newcommand{\geallevalbugs}{92}

\newcommand{\origspr}{40}
\newcommand{\sprevalbugs}{37}
\newcommand{\sprallevalbugs}{85}
\newcommand{\geTotSimilar}{8}
\newcommand{\geTotDel}{13}
\newcommand{\geallTotDel}{19}
\newcommand{\geTotLoc}{3}
\newcommand{\geTotSame}{16}
\newcommand{\geallTotSame}{26}
\newcommand{\geTotNoRep}{3}
\newcommand{\geTotSpeed}{0.95}
\newcommand{\geTotSpeedSame}{1.43}
\newcommand{\geallTotSpeedSame}{1.42}
\newcommand{\geTotCorrect}{3}
\newcommand{\geTotPla}{40}
\newcommand{\mgeTotCorrect}{2}
\newcommand{\mgeTotVeriYes}{38}
\newcommand{\mgeTotVeriNo}{3}
\newcommand{\mgeTotVeri}{(\mgeTotVeriYes, \mgeTotVeriNo)}
\newcommand{\sprTotCorrect}{12}
\newcommand{\sprTotPla}{24}
\newcommand{\msprTotCorrect}{13}
\newcommand{\msprTotVeriYes}{22}
\newcommand{\msprTotVeriNo}{1}
\newcommand{\msprTotVeri}{(\msprTotVeriYes,\msprTotVeriNo)}
\newcommand{\mgeTotSpace}{39}
\newcommand{\mgeallTotSpace}{40}
\newcommand{\msprTotSpace}{28}
\newcommand{\msprallTotSpace}{28}
\newcommand{\sprTotSimilar}{2}
\newcommand{\sprTotDel}{3}
\newcommand{\sprallTotDel}{10}
\newcommand{\sprTotLoc}{9}
\newcommand{\sprTotSame}{21}
\newcommand{\sprallTotSame}{42}
\newcommand{\sprTotNoRep}{1}
\newcommand{\sprTotSpeed}{1.66}
\newcommand{\sprTotSpeedSame}{1.25}
\newcommand{\sprallTotSpeedSame}{1.69}

\newcommand{\nodeleteallTotSpeedSame}{1.20}

\newcommand{\ourfix}{23}
\newcommand{\theirfix}{five}
\newcommand{\ourfixper}{65.7\%}
\newcommand{\theirfixper}{14.3\%}
\newcommand{\ourrrper}{8.3\%}
\newcommand{\theirrrper}{100\%}
\newcommand{\squeezeup}{\vspace{-0.5mm}}
\newcommand{\n}{\textit{anti-patterns}}
\newcommand{\vima}{f80e67~\cite{vim73202}}
\newcommand{\vimabug}{~\cite{vim73202bug}}
\newcommand{\vimb}{509890~\cite{vim509890}}
\newcommand{\vimbbug}{~\cite{vim509890bug}}
\newcommand{\vimc}{a3552c~\cite{vima3552c}}
\newcommand{\vimcbug}{~\cite{vima3552cbug}}
\newcommand{\vimd}{220906~\cite{vim220906}}
\newcommand{\vimdbug}{~\cite{vim220906bug}}
\newcommand{\indenta}{2.2.10~\cite{indenta}}
\newcommand{\indentabug}{~\cite{indentabug}}
\newcommand{\tara}{1.14~\cite{tara}}
\newcommand{\tarabug}{~\cite{tarabug}}
\newcommand{\pythona}{b878df~\cite{pythona}}
\newcommand{\pythonabug}{~\cite{pythonabug}}
\newcommand{\pythonb}{5b0fda~\cite{pythonb}}
\newcommand{\pythonbbug}{~\cite{pythonbbug}}
\newcommand{\perla}{dca606~\cite{perla}}
\newcommand{\perlabug}{~\cite{perlabug}}
\newcommand{\perlb}{bb9ee97~\cite{perlb}}
\newcommand{\perlbbug}{~\cite{perlbbug}}
\newcommand \cmmnt[1]{\textcolor{red}#1}

\newcommand{\analyzedPRs}{382}
\newcommand{\analyzedRepos}{296}
\newcommand{\analyzedMotivationPRs}{79}
\newcommand{\numberofmaincategories}{seven}
\newcommand{\numberofsubcategories}{26}
\newcommand{\numberofmotivations}{four}
\newcommand{\readabilitynumber}{21}
\newcommand{\readabilityratio}{26.58}
\newcommand{\complexitynumber}{7}
\newcommand{\complexityratio}{8.86}
\newcommand{\complexityclosetoratio}{9}
\newcommand{\deletionnumber}{49}
\newcommand{\deletionratio}{62.03}
\newcommand{\deletionapproximateratio}{60}
\newcommand{\reusabilitynumber}{3}
\newcommand{\reusabilityratio}{3.80}

\newcommand{\numberofnotrefactoringcategories}{eight}

\newcommand{\controllogicrelatednumber}{151}
\newcommand{\controllogicrelatedratio}{39.53}
\newcommand{\controllogicrelatedsubcategoriesnumber}{ten}
\newcommand{\functionreturnnumber}{53}
\newcommand{\functionreturnratio}{13.87}
\newcommand{\conditionalexpressionnumber}{32}
\newcommand{\conditionalexpressionratio}{8.42}
\newcommand{\foreachnumber}{15}
\newcommand{\foreachratio}{3.93}
\newcommand{\mergeconditionnumber}{13}
\newcommand{\mergeconditionratio}{3.40}
\newcommand{\ternarynumber}{11}
\newcommand{\ternaryratio}{2.88}
\newcommand{\pipelinenumber}{6}
\newcommand{\pipelineratio}{1.57}
\newcommand{\mergecatchnumber}{5}
\newcommand{\mergecatchratio}{1.31}
\newcommand{\reformatifelsenumber}{9}
\newcommand{\reformatifelseratio}{2.36}
\newcommand{\extractionnumber}{120}
\newcommand{\extractionratio}{31.41}
\newcommand{\extractmethodnumber}{75}
\newcommand{\extractmethodratio}{19.63}
\newcommand{\extractvariablenumber}{40}
\newcommand{\extractvariableratio}{10.47}
\newcommand{\apinumber}{62}
\newcommand{\apiratio}{16.23}

\newcommand{\diamondnumber}{16}
\newcommand{\diamondratio}{4.19}

\newcommand{\trywithnumber}{2}
\newcommand{\trywithratio}{0.52}

\newcommand{\notdetectedlanguagefeatureratio}{13.09}
\newcommand{\totallanguagefeatureratio}{25.92}

\newcommand{\bigdatasetunfilterreponumber}{26110}
\newcommand{\bigdatasetreponumber}{25022}
\newcommand{\bigdatasetcommitnumber}{37846}

\newcommand{\validaveragetestnumber}{96}
\newcommand{\validaveragemethodsize}{21}

\newcommand{\wholeaverageoriginalmethodsize}{17}
\newcommand{\wholeaveragechangedmethodsize}{14}

\newcommand{\validaverageoriginalmethodsize}{16}
\newcommand{\validaveragechangedmethodsize}{12}

\newcommand{\unchangeditemratio}{30}

\begin{abstract}
Several code generation models have been proposed to help reduce time and effort in solving software-related tasks. To ensure responsible AI, there are growing interests over various ethical issues (e.g., unclear licensing, privacy, fairness, and environment impact). These studies have the overarching goal of ensuring ethically sourced generation, which has gained growing attention in speech synthesis and image generation. In this paper, we introduce the novel notion of \emph{Ethically Sourced Code Generation} (\tooln) to refer to managing all processes involved in code generation model development from data collection to post-deployment via ethical and sustainable practices. To build a taxonomy of \tooln{}, we perform a two-phase literature review where we reviewed \manualread{} papers across various domains and specific to AI-based code generation. We identified \relevantpaper{} relevant papers with 10 initial dimensions of \tooln. To refine our dimensions and gain insights on consequences of \tooln, we surveyed \numreplyvalid{} practitioners, which include six developers who submitted GitHub issues to opt-out from the Stack dataset (these \emph{impacted users} have real-world experience of ethically sourced issues in code generation models). The results lead to 11 dimensions of \tooln{} with a new dimension on code quality as practitioners have noted its importance. We also identified consequences, artifacts, and stages relevant to \tooln{}. Our post-survey reflection showed that most practitioners tended to ignore society-related dimensions despite their importance. Most practitioners either agreed or strongly agreed that our survey help improve their understanding of \tooln{}. Our study calls for attention of various ethical issues towards \tooln. 
\end{abstract}

\maketitle
\renewcommand{\shortauthors}{Xu et al.}






\section{Introduction}\label{sec:intro}

Due to the rising demand in automated code generation to reduce software development and maintenance costs, companies like Meta and OpenAI have proposed several code generation models (e.g., Code Llama and Codex) for solving software engineering tasks (e.g., automated program repair~\cite{le2019automated,aprsurvey,lyu2024automatic} 
and program synthesis~\cite{googlellmsynthesis}).
However, with the advent of responsible AI, there are rising global concerns about various ethical dimensions of AI models to ensure ethical and sustainable development. 
First, practitioners are hesitant to use code generation models due to unclear licensing information and potential copyright infringements.  
For example, an Intercept lawsuit~\cite{interceptlaw} claimed that ``According to the award-winning website Copyleaks,
nearly 60\% of the responses provided by Defendants’ GPT-3.5 product in a study conducted by
Copyleaks contained some form of plagiarized content, and over 45\% contained text that was
identical to pre-existing content'', indicating the severity of copyright infringements in generated code. Second, several companies (e.g. Apple~\cite{appleban} and Samsung~\cite{samsung2023chatgpt})
have banned the usage of AI-based code generation models due to privacy concerns over data leakage. 
Third, several ethical guidelines~\cite{jobin2019global,hagendorff2020ethics} were introduced to govern and guide the design of responsible AI models by emphasizing various ethical principles (e.g., transparency, justice and fairness, non-maleficence, responsibility, and privacy).
Fourth, there exist emerging needs to consider the environmental impact of AI by characterizing the carbon footprint of AI computing ~\cite{wu2022sustainable,schwartz2020green,vartziotis2024learn}. With the overarching goal of ensuring \emph{ethically-sourced} generation, these initiatives encompass various ethical dimensions that practitioners need to consider when developing and adopting code generation models for software development and maintenance.  

``Ethically sourced'' refers to
``managing all processes of supplying the firm with required materials
and services from a set of suppliers in an ethical and
socially responsible manner''~\cite{kim2018ethical}.
In a broader view, ethical sourcing is usually associated with the concept \emph{triple bottom line}
that stresses the importance of measuring success in three key factors: people (social impact), planet (environmental impact), and profit (financial performance)~\cite{elkington1998partnerships}. Meanwhile, the core of ethical production involves \emph{supply chain management}, ensuring that each point of the supply chain is obtained ethically and sustainably. In the context of AI systems, the idea of AIBOM (AI Bills of Material)~\cite{xia2023empirical,xia2024trust} were introduced to describe all the components of an AI system to ensure AI transparency~\cite{Wang2024}.

The concept of ethical sourcing has been adopted in various domains (e.g., economy~\cite{peluso2023navigating} and medicine~\cite{kim2021korea}) but is still a new concept for the generative AI community. For speech synthesis~\cite{scott2019owns}, the ethical questions on voice donation (e.g., voice ownership and identity) were discussed. For image generation, Adobe released Firefly, an image generator, in March 2023~\cite{ethicalart}, claiming to use solely legal and ethical sources for training, making it less likely to induce copyright issues, and ``commercially safe''. However, there remains controversial debates on its ethical claim as a Bloomberg report~\cite{adobetrain} revealed that 5\% of its training data were not being ethically sourced as they were from 
 another image generator. As CodeGen models are at the early stage of adoption, it is important and timely to study the ethically sourced issues for code generation.

 \begin{figure}[tbp]
\includegraphics[width=1\linewidth]{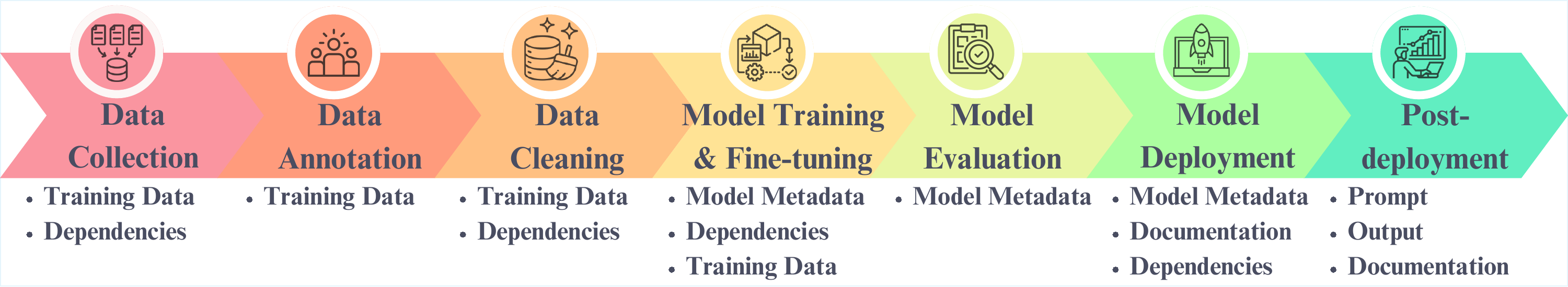}   
\caption{Supply Chain of \tooln (Stages and Artifacts)
}
\label{fig:sc}
\end{figure}

In this paper, to the best of our knowledge, we for the first time introduce the notion of \emph{Ethically Sourced Code Generation} (\tooln) to refer to managing all processes involved in code generation model development from data collection to post-deployment and artifacts involved via ethical and sustainable practices throughout the supply chain as shown in Figure \ref{fig:sc}. 
Prior ethically sourced research~\cite{kim2018ethical} revealed four main research areas that were widely discussed (i.e., socially responsible
 supply management, unethical purchasing behaviour, ethical sourcing codes of conduct,
 fair and ethical trade). Do software practitioners also care about the same set of dimensions when defining \tooln? Considering the unique characteristics of AI-driven code generation pipeline~\cite{wu2022sustainable}, what are the key software artifacts and stages that contribute to ethical code generation supply chain management? Prior studies show that unethical sourcing may lead to negative impacts such as loss of reputation~\cite{kim2022shareholder} and legal risks \cite{scott2019owns}.  Are these impacts any different in the case of unethically sourced code generation (Un\tooln)?  
 As the first study to understand the various dimensions of \tooln and consequences of Un\tooln, our study aims to answer the questions below: 
\begin{description}[noitemsep,topsep=0pt]
 \item[RQ1:] What are the dimensions that define \tooln{} from the practitioners' perspectives?
 \item[RQ2:] Which stages and what kinds of artifacts constitute an ethical supply chain management in the context of code generation? \item[RQ3:] What are the potential consequences of unethically sourced code generation (Un\tooln)?
 \item[RQ4:] What are the acceptable trade-offs and contamination when enforcing \tooln? 
\end{description}

Our study aims to raise awareness by proposing the concept and contributing to the body of knowledge about \tooln{}.
To understand the scope and various dimensions of ``ethically sourced'', we first conduct a systematic literature review of prior studies on it. Our interdisciplinary literature review covers publications across various domains, and lays the foundation towards a taxonomy of dimensions involved in defining \tooln{}. 

Based on our literature review, we conduct a survey questionnaire involving \numreplyvalid{} practitioners.  Specifically, the surveys targeted
software practitioners familiar with code generation, contributors of code generation models, and researchers in code generation. Our study expands our understanding of the various dimensions that define \tooln{}, and potential consequences of Un\tooln, identifies areas that research and
practice on \tooln{} support should focus on, and provides a
discussion of potential solutions to overcome these consequences.

In summary, this paper makes the following contributions:
\begin{description}[leftmargin=*,noitemsep,topsep=0pt]

        \item[Systematic Literature Review:] We perform a two-phase systematic literature review using a combination of database search, forward and backward snowballing to adopt a multidimensional view in defining \emph{ethically sourced code generation (\tooln)}. We first identified various dimensions of ethical sourcing from various domains, and then performed a domain-specific literature search.  We manually reviewed \manualread{} papers and identified \relevantpaper{} relevant papers. Our review identified 10 dimensions that define \tooln. 
         
        \item[Survey:] We surveyed \numreplyvalid{} practitioners to shed light on practitioners' views on \tooln{}, including their definitions of \tooln, evaluation on provided dimensions, the applicability of dimensions to stages and artifacts in the supply chain of \tooln{}, the dimensions that they tend to ignore in defining \tooln, potential consequences of Un\tooln, and acceptable trade-offs.
        \item[Discussion:] We discuss the gaps within existing code generation models and \tooln models, and highlight the key areas that researchers and practitioners should focus on to build \tooln{}.
        

        \item[Evaluation:] We also evaluated the usefulness of our survey. In the post-survey reflection, most participants either agreed (56.3\%) or strongly agreed (18.8\%) that their understanding of \tooln{} has improved after taking our survey, indicating that our survey helped raise awareness of the importance of \tooln. 
        Our anonymized survey questionnaire and results are publicly available at: \url{https://github.com/xuzhi2021/ESCodeGen}.
\end{description}
In summary, we noted the key findings below from our study:
\begin{itemize}[leftmargin=*,noitemsep,topsep=0pt]
\item \tooln{} is defined by 11 important dimensions: (1) Subject Rights, (2) Equity, (3) Access, (4) Accountability, (5) Intellectual Property (IP) Rights, (6) Integrity, (7) Code Quality, (8) Social Responsibility, (9) Social Acceptability, (10) Labor Rights, (11) Environmental Sustainability. Among these dimensions, stakeholders were most concerned about (1), (5), and (11), and tended to ignore (8), (9), and (10) despite their importance.
\item Developers who spent time opting-out from the training data of the Stack dataset~\cite{bigcode_optout_v2} (\emph{impacted users}) noted the importance of opt-in consent as a key dimension of \tooln.
\item All stages and artifacts in \tooln's supply chain are considered to be relevant to the dimensions.
\item Participants have listed many consequences of Un\tooln, e.g., exploitation of developers, low-quality code, monopolization of generative AI, and impact on open-source communities. Among the given options, most stakeholders cared about lawsuit issues, security, and privacy risks.
\item When considering various trade-offs in ensuring \tooln, most participants noted that accuracy loss is unacceptable (most can accept up to 10\% accuracy loss). 
\item Most participants thought that none of the existing code generation (CodeGen) models aligns with the definition of
\tooln or only partially aligns (need to improve on transparency and opt-in consent), indicating a considerable
gap between current models and \tooln.
\end{itemize}


\section{Background and Related Work}

We will discuss most of relevant papers in our literature review in Section~\ref{sec:review}. In this section, we focus on two lines of research closely related to our work: (1) approaches related to \tooln, (2) empirical studies on ethically sourced issues.  

\noindent\textbf{Ethically Sourced Code Generation (\tooln).} Recent trends in open-source code generation models focus on various dimensions towards ethical generation, including enhancing AI transparency~\cite{bigcode2024stackv2dedup,olmo20242}, green code generation~\cite{tuttle2024can, vartziotis2024learn}, bias evaluation of code generation (CodeGen)~\cite{feng2023investigating, liu2024toward}. 
To enhance transparency of training data, the Stack dataset~\cite{bigcode2024stackv2dedup} contains source code from permissive licenses, and its data governance plan allows developers who do not wish to have their code used for pretraining large language models to opt-out. The Stack dataset is used to train many code generation models (e.g., StarCoder~\cite{li2023starcoder} and OLMo2~\cite{olmo20242}). 

\noindent\textbf{Studies on \tooln.}
Most studies focus on understanding ethically sourced issues in various domains~\cite{kim2018ethical,boina2023balancing}, and some focus on the software supply chain~\cite{xia2023empirical,xia2024trust}. The most closely related work to our work is the literature review of ethical sourcing~\cite{kim2018ethical} (published in July 2018),  which combines a systematic literature review and a citation network analysis to understand the relationship between the reviewed literature. Although we rely on a literature review to derive the dimensions defining \tooln{}, our work differs in that: (1) instead of citation network analysis, we perform a two-phrase hierarchical review to obtain diverse perspectives of ethically sourced dimensions, (2) our focus on code generation offers a technical view of ethically sourced issues, which is absent in prior work, (3) compared to prior study published on 2018, our study offers an updated view of ethically sourced issues.    
\section{Systematic Literature Review}
\label{sec:review}

To obtain a comprehensive definition of \tooln, we conducted a two-phase hierarchical systematic literature review (broad $\rightarrow$specific), using a combination of database searching, forward and backward snowballing~\cite{jalali2012systematic}. Figure~\ref{fig:overall} shows the overall workflow of our literature review. 
To obtain a broad and high-level view of ethical sourcing from multiple disciplines, we first conducted a survey of research across various domains
(e.g., economy~\cite{peluso2023navigating}, medicine~\cite{kim2021korea}, and religion~\cite{alzeer2025halal}) as there is a lack of a clear definition of \tooln.  
In the second phase, we focused specifically on LLM-based code generation literature and annotated the relevant dimensions to the development of \tooln. 

\subsection{Cross-disciplinary General Literature Search}
\subsubsection{Initial Database Search}
We searched for ``ethically sourced'' on Google Scholar to obtain a comprehensive definition and related dimensions across various domains. Since this term is not specific to the computer science field, we did not use specialized databases such as IEEE Xplore or ACM Digital Library. The search returned about 112,000 results, which were too many to review in full, so we manually examined the top 100 papers sorted by relevance. After excluding non-scholarly publications (55 papers remaining) and selecting relevant ones, we obtained 45 papers. Among the top 100, we found that after the top 51 papers, no new dimensions emerged. Based on this review, we established 
the synonyms of ``ethically sourced'' as the first set of keywords for the domain-specific literature review. 

\begin{figure}[tbp]
    \centering
    \includegraphics[width=1\linewidth]{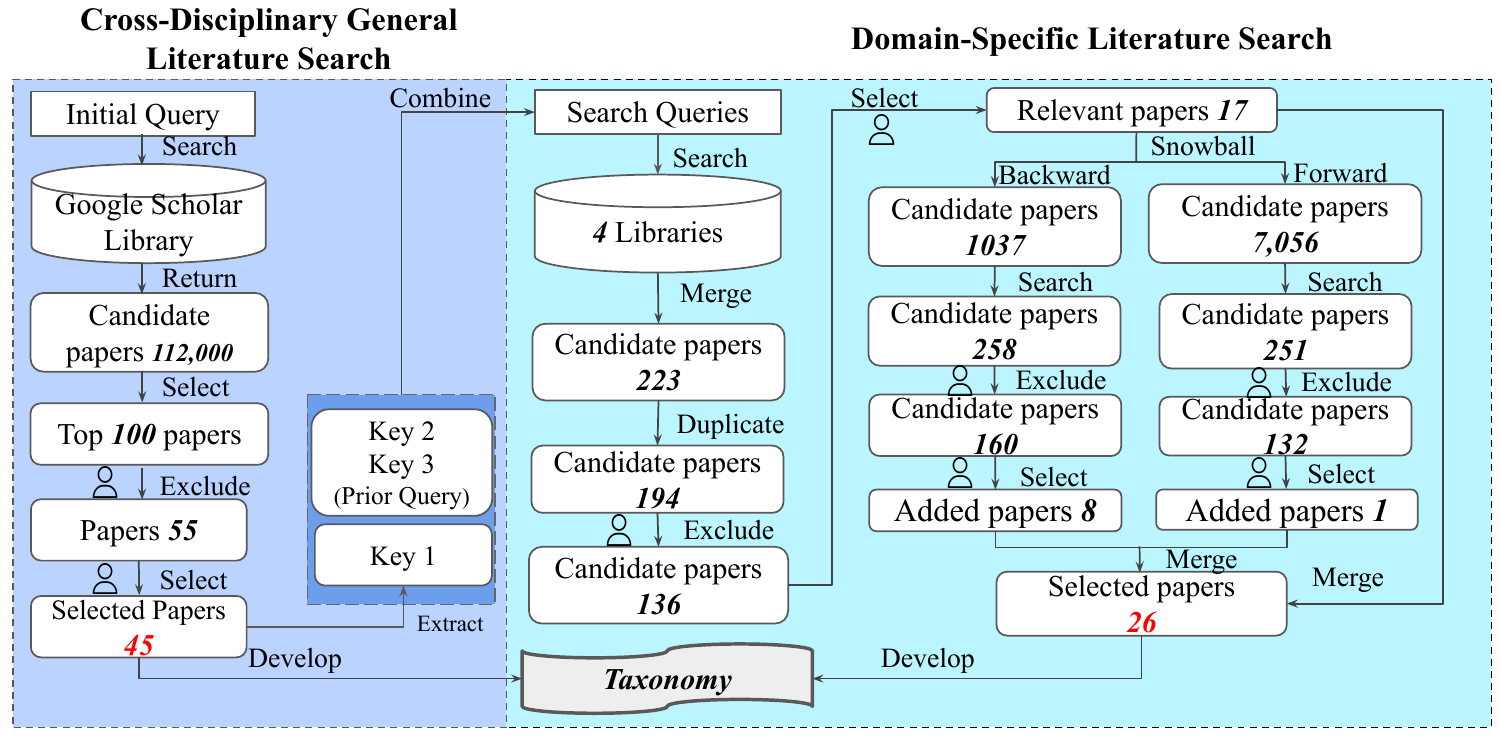}
     \caption{The Workflow of Our Literature Review}
    \label{fig:overall}
\end{figure}

\subsubsection{Constructing Keywords for Queries}
We prepared three sets of keywords ($key1$, $key2$, $key3$) to design the queries for the domain-specific literature review. From the general literature search, we collected $key1$, synonyms of ``ethically sourced''.
Then, we reused the derived synonyms of ``artificial intelligence'' and ``code generation'' from prior work \cite{negri2024systematic} as $key2$ and $key3$. Our summarized keyword sets are as follows:


\begin{description}[leftmargin=*]
  \item[$key1$ :] ethically sourced, ethical sourcing, ethically acquired, ethically procured, ethical acquisition, responsible sourcing, responsibly sourced, sustainable sourcing
  \item[$key1$':] ethic* OR responsibl* OR sustainabl*
  \item[$key2$:] LLM, AI, artificial intelligence, LLMs, large language models
  \item[$key3$:] code generation, code creation, generate code, code writing, code production, code correction, code quality
\end{description}

\begin{table}[tbp]
\centering
\small
\caption{Database Search Queries and Scopes}
\label{tab:query}
\resizebox{\linewidth}{!}{%
\begin{tabular}{l|l|l}
\hline
Database       & Query Keywords Summary                                                            & Search Scope                      \\ \hline
IEEE           & $key1$' AND $key2$ AND $key3$                                            & All Metadata                      \\
ACM            & $key1$' AND $key2$ AND $key3$                                            & Abstract                          \\
Scopus         & $key1$' AND $key2$ AND $key3$                                            & Article Title, Abstract, Keywords \\
Google Scholar & $key1$ * $key2$ * $key3$  & Anywhere (Default)                           \\ \hline
\end{tabular}
}
\end{table}


\subsection{Domain-Specific Literature Search}

\subsubsection{Query Design and Database Search} 
Table \ref{tab:query} shows the queries and scopes for four popular databases in the field of computer science: IEEE, ACM, Scopus, and Google Scholar. Initially, we found that queries in the form of $key1$ AND $key2$ AND $key3$ returned zero results in IEEE, ACM, and Scopus, so we simplified and broadened $key1$ to $key1$' using wildcard operators. For example, we used ``ethic*'' to cover multiple forms such as ``ethics'', ``ethical'', and ``ethically''. 
As the length of our query (362 characters without spaces) exceeded the limit for Google Scholar (256 characters), and Google Scholar does not support wildcard operators, we searched each combination of $key1$, $key2$, and $key3$ individually, then collected the results, merged and deduplicated them. 
Since the three databases provide different scope options, we chose a different but appropriate scope for each to include as many results as possible. 
The query in Google Scholar was finished on March 12, 2025. The queries in IEEE, ACM, and SCOPUS were completed on March 14, 2025.



\subsubsection{Merging, Deduplication and Exclusion} We obtained 61 papers from Google Scholar, 79 from IEEE Xplore, 19 from ACM Digital Library, and 64 from Scopus individually, leading to a total of 223 papers. After merging the list of papers and deduplication, we obtained 194 papers.
Then, we manually excluded papers that are either not in English or not published in scholarly journals or conferences, and eventually obtained 136 papers.

\subsubsection{Paper Selection}
As shown in Figure~\ref{fig:overall}, we performed paper selection in several rounds (on 55, 136, 160, and 132 papers). Specifically, we manually excluded irrelevant papers that 1) only reflect on the keywords in sections like ``Discussion'' or ``Related Work'' instead of discussing the keyword itself, and 2) superficially mention the keywords, but do not focus on ethical sourcing issues.  

\subsubsection{Snowballing}
We conducted forward and backward snowballing on the 17 relevant papers. This process yielded 7,056 papers through forward snowballing and 1,037 papers through backward snowballing. 
We then applied keyword filtering to remove irrelevant papers, resulting in 251 and 258 papers, respectively. After manual exclusion, we retained 132 forward and 160 backward papers that were English academic publications. In the last step, we identified one additional relevant paper from forward snowballing and eight papers from backward snowballing.


Combined with the original 17 papers, this resulted in 26 relevant papers. 
In total, we manually reviewed 100 (general search) + 194 (specific search) + 251 (forward snowballing) + 258 (backward snowballing) = 803 papers, and got \relevantpaper (45+26) relevant papers. 

\subsection{Construction of the Taxonomy} 
\label{sec: taxonomy}

\begin{table*}[tbp]
\centering
\caption{Ethics Dimension Taxonomy for \tooln (refer to our website for the complete list of references)}
\small
\label{tab: taxonomy}
\makebox[\textwidth]{
\resizebox{1\textwidth}{!}{
\begin{tabular}{llllll}
\toprule
\textbf{Area} & \textbf{Dimension}    & \textbf{Aspect}            & \textbf{Description}                                                                                                                    & \textbf{Domain-specific Papers}                                                                                   & \textbf{General Papers}                                                                                                                       \\ \midrule
Source        & Subject Rights (D1)         & Informed Consent (Opt-in')          & Data subjects are fully informed and give consent to source usages& \cite{listarcoder, majeed2024reliability, parikh2023empowering, prather2023robots}& \cite{kim2021korea, gupta2022adima, belisle2023individuals, belisle2024stakeholder, belisle2025consent}\\
              &                       & Privacy and Security                   & Both inputs and outputs avoid private or confidential information& \cite{klemmer2024using, niu2023codexleaks, depalma2024exploring, garcia2025teaching, gupta2023chatgpt}& \cite{belisle2023individuals, belisle2024stakeholder, belisle2025consent}\\
              & Equity (D2)                & Diversity                  & Diverse perspectives (e.g., geographic) are considered                                              & \cite{chowdhery2023palm, listarcoder, majeed2024reliability, muthazhagu2024exploring, parikh2023empowering}& \cite{pacia2024early, fessenko2024ethical, goebel2012influence, gupta2022adima, kamstra2023novel}\\
              &                       & Fairness                   & Both inputs and outputs are free from biases                                                                                    & \cite{feng2023investigating, liu2024toward, ling2024evaluating, liu2023uncovering, aswal2024empowering}& \cite{fessenko2024ethical, goebel2012influence, gupta2022adima, nzalasse2024sig, pacia2024early}\\
              &                       & Representativeness             & Proportional coverage of different groups are considered                                             & \cite{becker2023programming, feng2023investigating, lau2023ban, muthazhagu2024exploring}& \cite{scott2019owns, belisle2025consent, rameau2024developing, pacia2024early, belisle2024stakeholder}\\
              & Access (D3)                & Accessibility              & Resources are available under appropriate conditions                                                                 & \cite{gupta2023chatgpt, listarcoder, prather2023robots}& \cite{rameau2024developing, belisle2025consent, gupta2022adima, kamstra2023novel, kim2021korea}\\
              & Accountability (D4)        & Transparency               & The inputs and development processes are visible                                         & \cite{ma2024code, majeed2024reliability, muthazhagu2024exploring, parikh2023empowering, wang2024time}& \cite{kim2022shareholder, sanghavi13sustainability, hamad2024facilitating, mboga2017ethical, pacia2024early}\\
              & IP Rights (D5)             & Source Acknowledgement& Sources are properly acknowledged and credited                                                                                  & \cite{klemmer2024using, li2023discriminating, ma2024code, majeed2024reliability}& None\\
              &                       & Licensing                  & The use of inputs complies with relevant licenses                                                                               & \cite{liu2024toward, becker2023programming, lau2023ban, ma2024code, majeed2024reliability}& \cite{kamstra2023novel, scott2019owns}\\
              &                       & Generation Distinctiveness& The outputs are sufficiently distinct from the origins                                & \cite{yan2022whygen, li2023discriminating, wang2024time, prather2023robots, ma2024code}& None\\
              & Integrity (D6)            & Contamination              & Inputs are verified and remain free from contamination            & 
None& \cite{alzeer2025halal, azmi2024combining}\\
              & Code Quality+ (D7)       & Accuracy+             & The generated code is accurate                                                                                                  & \cite{chowdhery2023palm, liu2024toward, majeed2024reliability, parikh2023empowering, wang2024time}& \cite{belisle2023individuals, belisle2024stakeholder}\\\hline
Society       & Social Responsibility (D8) & Community Development      & Communities involved in or affected  by the development benefit from it& 
None& \cite{ameta2024green, kim2022shareholder, mboga2017ethical, pulverail2013ethical, roumeliotis2023unveiling}\\
              & Social Acceptability (D9)  & Philosophy and Religion & Religious beliefs, practices, and sensitivities are respected                                                  & \cite{chowdhery2023palm, listarcoder, liu2023uncovering}& \cite{alzeer2025halal, azmi2024combining, billings2024leading, nadal2021charting}\\
              &                       & Culture                    & Diverse cultural identities and values are considered                                                   & 
\cite{chowdhery2023palm, majeed2024reliability}& \cite{billings2024leading, belisle2024stakeholder}\\
              & Labor Rights* (D10)          & Fair Wages*                 & Human workers involved (e.g., authors of training data*) receive fair pay                                  & \cite{lau2023ban}& \cite{adebayo2024sustainable, azmi2024combining, gupta2022adima, kim2018ethical, kim2022shareholder}\\
              &                       & Working Conditions         & Mentally and physically safe conditions are provided                                                                            & 
\cite{lau2023ban}& \cite{adebayo2024sustainable, barrientos2012ethical, goebel2012influence, kim2022shareholder, mboga2017ethical}\\
              &                       & Legal Employment           & The labor of  vulnerable communities is not exploited illegally& None& \cite{adebayo2024sustainable, goebel2012influence, kim2022shareholder, mboga2017ethical, peluso2023navigating}\\\hline
Enviro-& Sustainability (D11)   & Energy Consumption         & The amount of energy and resources used is considered& \cite{vartziotis2024learn, sikand2024generative, tuttle2024can, becker2023programming, chowdhery2023palm}& \cite{belisle2024stakeholder, belisle2025consent, blowfield2000ethical, goebel2012influence, holmsten2017luxury}\\
              nment&                       & Emissions and Pollution& The environmental pollution (e.g., carbon emissions) is considered& \cite{chowdhery2023palm, lau2023ban, sikand2024generative, vartziotis2024learn}& \cite{ameta2024green, blowfield2000ethical, goebel2012influence, holmsten2017luxury, mboga2017ethical}\\ \bottomrule
\end{tabular}
}
}

{\footnotesize
\begin{description}[leftmargin=*]
  \item[1] The ``Source'' area encompasses dimensions focusing on model and data governance.
  \item[2] The superscript + means that the dimension is a new technical dimension derived from the result of the survey.
  \item[3] The superscript * denotes that the refined dimension where survey participants noted compensation to authors or input training data is desirable. 
  \item[4] The superscript ' denotes that the refined aspect where survey participants noted that opt-in informed consent is preferred over opt-out.
 \end{description}
} 
\end{table*} 

We recorded candidate themes when reading the \relevantpaper{} relevant papers to build a taxonomy with various dimensions that define \tooln{}. Specifically, we followed the recommended steps of thematic analysis~\cite{cruzes2011recommended}, 
and assessed inter-rater disagreement at two key steps. 1) Extract data: Two authors (coders) independently extracted dimension-relevant excerpts 
for all these papers. The percentage agreement was nearly 90\%. A codebook defining various dimensions was also developed in this process and finalized through discussions. 2) Identify and code relevant dimensions: Both coders independently coded all extracted excerpts using the codebook. 
We transformed the multi-label coding task into a label-level binary matrix, treating each (excerpt, label) pair as a unit, and computed Cohen’s Kappa coefficient~\cite{cohen1960coefficient}, and average Jaccard Index~\cite{stemler2004comparison} per excerpt to reflect annotation set-level consistency. 
In initial coding (first 10\% of excerpts), the Cohen’s Kappa was 0.62 (moderate agreement). After a calibration meeting to refine code definitions, subsequent coding's Kappa value achieved 0.91 (almost perfect agreement), with the post-calibration average Jaccard Index 0.84 (relatively high similarity). Steps (3)-(5) involved discussions to translate codes into themes, create higher-order themes, and assess trustworthiness. These were performed via multiple meetings among the authors to ensure consistency and validity.


Table~\ref{tab: taxonomy} shows a structured taxonomy resulting from our thematic analysis, reflecting key dimensions of \tooln through three hierarchical levels: (1) \textbf{Area}, representing broad areas of ethical concern; (2) \textbf{Dimension}, grouping related issues within each dimension; and (3) \textbf{Aspect}, capturing concrete and specific ethical aspects, each with a brief description and relevant citations from the two-phase literature review. This structure allows for both high-level reasoning and fine-grained analysis of ethical sourcing practices in the development of \tooln. 

\subsubsection{Source}
This area encompasses the dimensions related to the source of \tooln{}, including subject rights, equity, access, accountability, intellectual property rights, and integrity.

\noindent\textbf{Subject Rights}. This refers to whether data subjects' rights on controlling the use of their source are respected, including:

\underline{\noindent\emph{Informed Consent.}} 
In most domains (e.g., biomedical field gathering stem cells \cite{kim2021korea}, and voice generation gathering voice\cite{gupta2022adima}), the standard mechanism for obtaining consent is opt-in, where data subjects are explicitly asked whether they agree to the use of their data for specific purposes. 
However, CodeGen models often rely on large-scale datasets mined from the web (e.g., GitHub), making obtaining individual consent impractical. To address this, developers of the Stack dataset~\cite{bigcode_optout_v2} introduced an opt-out mechanism to allow developers to check if their data has been included and to request its removal from the training. However, developers who chose to opt-out pointed out that the mechanism might still be illegal and morally wrong~\cite{opt-out-issue}. To obtain real-world experience of ethically sourced issues in CodeGen, we invited these developers (known as \emph{impacted users} in Section~\ref{recruitment}) to our survey. 

\underline{\noindent\emph{Privacy and Security.}} 
Prior work \cite{klemmer2024using} mentioned some famous examples like Samsung employees accidentally disclosing sensitive company data to ChatGPT~\cite{samsung2023chatgpt}, 
and companies like Google explicitly warning employees not to enter confidential information into chatbots \cite{dastin2023google}. 
To address this, some techniques like CodexLeaks \cite{niu2023codexleaks} are proposed to test privacy leaks of CodeGen models.





\noindent\textbf{Equity}. This refers to whether the model development reflects principles of diversity, fairness, and representativeness.

\underline{\noindent\emph{Diversity}}. 
Several studies~\cite{parikh2023empowering,majeed2024reliability,pacia2024early} highlighted the importance of diversity in the source of models. 
Diversity is also emphasized in domains such as medical AI, as prior work \cite{pacia2024early} noted that ethical AI development is approached through interdisciplinary collaboration and the early involvement of diverse teams across the AI lifecycle.


\underline{\noindent\emph{Fairness}}. 
Several studies revealed that the code generated by ChatGPT and Codex exhibits biases related to demographic factors and stereotypes based on gender, race, emotions, class, naming structures, and other characteristics \cite{feng2023investigating, liu2024toward}. Several techniques focus on evaluating the bias of CodeGen models \cite{ling2024evaluating, liu2023uncovering}.


\underline{\noindent\emph{Representativeness}}. 
Several studies discussed this aspect to avoid inequity \cite{scott2019owns, belisle2025consent, rameau2024developing, lau2023ban, pacia2024early}.
For example, for voice datasets collection, prior study \cite{belisle2024stakeholder} stressed the need for targeted recruitment and oversampling from underrepresented groups. For CodeGen, another work \cite{lau2023ban} stated that university programming instructors raised several equity-related concerns, including whether the training data behind AI tools sufficiently represents diverse groups.







\noindent\textbf{Access.} This refers to whether resources used in the development of models and models themselves are accessible or properly controlled.

\underline{\noindent\emph{Accessibility.} }
Recent work like StarCoder~\cite {listarcoder} stressed that closed-access models pose ethical challenges by preventing public oversight, hindering safety assessments, and limiting community contributions. Meanwhile, balancing between privacy and accessibility should also be considered when resources like training data contain sensitive or identifiable information~\cite{rameau2024developing}.
Moreover, ensuring that AI models are not prohibitively expensive or restricted to certain regions or populations (equitable access) is also important \cite{belisle2025consent}. 




\noindent\textbf{Accountability.} Source accountability refers to whether the development of CodeGen models ensures responsibility and traceability throughout the process, including:

\underline{\noindent\emph{Transparency.}} 
The importance of supply chain transparency is widely emphasized across domains \cite{kim2022shareholder, sanghavi13sustainability,hamad2024facilitating}. 
In the context of AI, developers or organizations should disclose what data is used and how outputs are produced to ensure input transparency and output explainability 
\cite{wang2024time}. Several regulatory frameworks (e.g., the GDPR for data processing \footnote{{GDPR}: Principles relating to processing of personal data (2023). Available at: \url{https://gdpr-info.eu/art-5-gdpr/}. Accessed: 2025-06-04}, and the European Commission’s AI Act for deployment 
\footnote{European Commission (2021). Proposal for a Regulation laying down harmonised rules on artificial intelligence (Artificial Intelligence Act). Available at: \url{https://eur-lex.europa.eu/legal-content/EN/TXT/?uri=CELEX\%3A52021PC0206}. Accessed: 2025-06-04}) 
stressed the importance of transparency across different stages of AI systems. Moreover, transparency regarding data sources is important in preventing the misuse of protected or copyrighted content \cite{majeed2024reliability}. 







\noindent\textbf{Intellectual Property (IP) Rights.} This concerns whether the development of CodeGen models respects IP rights related to their inputs.

\underline{\noindent\emph{Source Acknowledgement.}} 
Although related to transparency, this aspect specifically emphasized the responsibility to provide clear attribution, allowing users to trace the origin of code components to enhance traceability \cite{majeed2024reliability, li2023discriminating}. 



\underline{\noindent\emph{Licensing.}} 
Although many open-source repositories are accessible, many come with specific license terms on permissible uses (e.g., commercial use). However, studies have highlighted that many models have been trained on such repositories without adequately considering or adhering to these licensing terms, raising concerns about legal violations \cite{liu2024toward, becker2023programming, lau2023ban, ma2024code,majeed2024reliability}. For example, the GitHub Copilot investigation raised concerns on potential violation of open-source licenses on AI-generated code 
\cite{githubcopilotinvestigation}.







\underline{\noindent\emph{Generation Distinctiveness}.} 
Since models can learn and memorize details from the training data, there is a risk that generated outputs may closely mimic or even replicate code in training dataset, raising concerns about compliance and infringement \cite{yan2022whygen}. Hence, developers should ensure generated code maintains adequate distinctiveness from code in training data \cite{li2023discriminating, wang2024time}. Otherwise, the users who deploy generated code that is too similar to its source may unintentionally commit plagiarism \cite{prather2023robots}. 


\noindent\textbf{Source Integrity.} This considers whether all inputs are sourced from verified sources that comply with ethical standards, with no contamination involved during model development.

\underline{\noindent\emph{Contamination.}} 
This aspect is inspired by studies on ingredient evaluation and Halal certification~\cite{alzeer2025halal, azmi2024combining}, which stated the need of source raw materials strictly from certified suppliers to prevent cross-contamination from non-Halal ingredients. 




\subsubsection{Society}
The social area encompasses social responsibility, social acceptability, and labor rights.

\noindent\textbf{Social Responsibility}. This concerns whether the organization or developers of CodeGen models take on broader social responsibilities \cite{kim2022shareholder, ameta2024green}, including:


\underline{\noindent\emph{Community Development}.} This refers to whether communities involved in or affected by the development of CodeGen models benefit from it, rather than being treated solely as sources of data or labor. For example, Starbucks showed community commitment by adhering to fair trade principles, collaborating with coffee-producing communities, and organizing initiatives (e.g., donations)~\cite{kim2022shareholder}. 




\noindent\textbf{Social Acceptability}. This refers to if the development process of CodeGen models respects different cultures and religious beliefs.

\underline{\noindent\emph{Philosophy and Religion}.} This aspect concerns whether philosophical concerns, religious beliefs and sensitivities are respected during model development. 
For example, prior work \cite{nadal2021charting} discussed ethical concerns raised by certain religious communities regarding COVID-19 antibody vaccines that utilize reagents derived from aborted fetal cell lines. 
Similarly, for CodeGen, developers should ensure that inputs, outputs, and overall development processes align with commonly held religious principles and avoid content that conflicts with religious sensitivities. 

\underline{\noindent\emph{Culture}.} 
For example, body donor programs have emphasized integrating cultural and religious needs into operational practices \cite{billings2024leading}. Prior study \cite{belisle2024stakeholder} also mentioned that trustworthy voice AI should respect cultural sensitivity throughout data collection, synthetic data generation, and implementation. These principles about culture are equally applicable to CodeGen models.

    
    


\noindent\textbf{Labor Rights}. This dimension concerns whether organizations or developers respect the basic labor and human rights of workers involved in the development processes of CodeGen models.

\underline{\noindent\emph{Fair Wages}.} 
For example, a prior study \cite{lau2023ban} cited a news\footnote{\label{fn:openai_kenya}Billy Perrigo (2023). OpenAI Used Kenyan Workers on Less Than \$2 Per Hour To Make ChatGPT Less Toxic. \emph{Time}. Available at: \url{https://time.com/6247678/openai-chatgpt-kenya-workers/}. Accessed: 2025-05-07.} that OpenAI underpaid the Kenyan data labeler (less than \$2/hour), raising concerns about fair pay in the global AI supply chains. 

\underline{\noindent\emph{Working Conditions}.} 
For example, reducing a model’s toxicity often requires human labelers to handle harmful content, which may affect their mental health. In the previous news\textsuperscript{\ref{fn:openai_kenya}}, workers reported that the wellness services provided by the organization were limited and ineffective under the strain of heavy workloads.

\underline{\noindent\emph{Legal Employment}.} 
For example, previous research \cite{kim2022shareholder} cited a report \cite{zadek2007path} revealing child labor in global brands such as Nike, which demonstrated the importance of this aspect. 

    
    


\subsubsection{Environment} This area examines whether the development of CodeGen models aligns with principles of sustainability.

\noindent\textbf{Sustainability.} This dimension assesses whether the development of models is conducted in an environmentally sustainable manner.

\underline{\noindent\emph{Energy Consumption.}} 
Several studies \cite{sikand2024generative, vartziotis2024learn} have raised concerns about the high energy cost of LLMs and proposed approaches to evaluate energy efficiency (e.g., data center energy efficiency, hardware efficiency, and algorithmic efficiency). However, the generated code of existing models often overlooks basic energy-saving practices~\cite{sikand2024generative}, and typically results in higher runtime and energy consumption compared to human-written solutions~\cite{tuttle2024can}.

\underline{\noindent\emph{Emissions and Pollution.}} 
Several studies noted the concerns on this dimension and discussed about the solutions~\cite{chowdhery2023palm, lau2023ban, sikand2024generative}. For example, the organization can reduce the carbon emissions by locating the data center in regions with cleaner energy grids or by sourcing electricity from low-carbon or renewable providers \cite{vartziotis2024learn}.


    



\noindent\textbf{Excluded Dimensions.} We exclude some dimensions that are inapplicable to the field of CodeGen: biodiversity conservation under the environment dimension\cite{peluso2023navigating, ameta2024green}, animal ethics (animal welfare) \cite{kishore2021comparative, mitchell2024investigating}, health and safety (in medicine or chemistry) \cite{kim2021korea, ameta2024green}, and fair trade (in economy)\cite{kim2022shareholder, sanghavi13sustainability}. 

\begin{mybox}
\textbf{Finding 1:} Our literature review revealed that \tooln are defined by 10 dimensions from 3 areas, including 19 aspects. 
\end{mybox}


\section{Ethical Considerations} The research ethics committee of our institution requires that each researcher doing research involving human subjects obtain certification on a research ethics course (i.e., an online course with tutorials and exercises on ethics guidance) so all researchers involved have completed the course before getting ethical approval. Then, we submitted the methodology for the study (including the procedures to gather contact information, recruitment materials,
 survey questions), and received ethical approval from the research ethics committee of the institution directly involved in the survey. 

 At the start of the survey, we obtained informed consent from participants. We stressed that this is an anonymous survey, participation is voluntary, and they can quit the survey at any time, and incomplete responses will not be recorded. As fair labor is one key dimension in ethical sourcing, we provide five Amazon gift cards valued at 50 (in our designated currency) for every 50 participants as compensation, complying with the minimum wage requirement in our region. To protect the participants' anonymity, 
 we stressed that participation in the lucky draw is optional, and we collected their email addresses to distribute the gift cards in a separate, independent questionnaire.  
 
\section{Survey Design and Results Analysis}

Building upon the dimensions that we obtained via our literature review, we designed a questionnaire to elicit developers’ perspectives on whether these dimensions are appropriate for defining ``ethically sourced code generation,'' and to understand their underlying reasons and concerns. In addition, we investigated the broader significance of \tooln, including scenarios in which such ethical considerations are most critical. 

\subsection{Participants Recruitment}
\label{recruitment}
Our survey targeted stakeholders involved in the research and the development of CodeGen models. To identify these stakeholders, we invited participants via several rounds of invitations: (1) we shared the link of our survey through social media posts, (2) we collected a list of emails of authors of the relevant papers in our literature review (we only kept authors relevant to LLM research and ethical issues), leading to a total of 831 emails (out of these emails, we received six replies saying that they would not participate), (3) we contacted researchers within our social network and sent emails to mailing lists for software engineering and AI researchers. 

\noindent \textbf{Impacted Users.} In the last round of invitations, we included GitHub users who have opted-out from the Stack dataset~\cite{bigcode_optout_v2}, as they were affected by ethically sourced issues. Our email started with \emph{``As you requested to opt-out your repositories from The Stack dataset, we would like to invite...''}. As these users have experience of their repository being used as training data and chose to opt-out, their opinions are invaluable (we call them \emph{impacted users}). Interestingly, we observe that the response rate for impacted users is relatively high, 
i.e., we obtain six new responses in our last round of invitations out of 50 emails sent (12.0\% response rate compared to 3.3\% for previous invitations), showing that \emph{impacted users are more likely to share their views on \tooln}.

As a targeted survey on CodeGen, we listed several inclusion criteria at the start of the survey and in our invitation email: (1) adults aged 18 and over, (2) have at least one year of programming experience, (3) have at least some familiarities with CodeGen models (as a user, developer, or researcher), (4) not affiliated with the research group conducting the survey (to avoid conflict of interest). Finally, we received \numreply{} survey responses, of which \numreplyvalid{} were complete and valid (i.e., we analyzed the results from the \numreplyvalid{} ones).

\subsection{Survey Design} We designed the survey questionnaires based on our literature review, general guidelines for survey design~\cite{groves2011survey}, and SE-specific guidelines~\cite{kitchenham2002principlepart,kitchenham2002principles2}. 
The survey includes five types of questions: (1) short-answer open-ended questions,  (2) multiple-choice questions, (3) rating questions (with 5-point Likert scale), (4) matrix table questions, and (5) ranking questions. 
The survey consists of the six sections below: 


\noindent\textbf{1. Background.} This consists of five multiple choice questions and three open-ended questions, asking about the participant's non-identifiable background information. Our goal is to check whether participants are familiar with programming, CodeGen models, and the concept of ``ethically sourced''.

\noindent\textbf{2. Open-ended Questions on Defining \tooln.}
The second section investigated participants' opinions on what should be considered when defining ``ethically sourced code generation'' and whether they have ever encountered ethical sourcing issues through two open-ended questions. To minimize potential bias and to allow participants to provide the definition in their own words, we posed these open-ended questions prior to presenting our collected dimensions, thereby reducing the likelihood that participants would be influenced by our taxonomy.

\noindent\textbf{3. Dimension Evaluation.}
Here we first described each dimension and asked about the relevance of each to the definition of \tooln through 10 rating questions, with options rated in a five-point Likert scale (from``Not relevant (1)'', to ``Very relevant(5)''). 
Secondly, we asked which development stages and artifacts (columns) in the development of CodeGen models they believe each dimension (rows) applies to, with four matrix table questions. 

\noindent\textbf{4. Real-World Experiences and Reflections.}
This part asked about participants' personal reasons for rating the dimensions, their real-world experiences or thoughts related to \tooln, and potential impacts of Un\tooln by four open-ended questions and one multiple choice question. 

\noindent\textbf{5. Trade-offs and Acceptable Costs.}
This section has five multiple-choice questions on respondents' views on the acceptable costs of making a CodeGen model ethically sourced, and one ranking question on the relative importance of each trade-off option.

\noindent\textbf{6. Post-survey Reflections.}
The final section asked participants to reflect on their experience with this survey. For example, whether they think that \tooln{} is more important than they thought after taking the survey (from``Strongly disagree (1)'', to ``Strongly agree(5)''), which dimensions were ignored by them before taking this survey (by a multiple choice question), and whether the dimensions are generalizable to other fields beyond CodeGen (e.g., voice generation) (from``Definitely not (1)'', to ``Definitely yes(5)''). This part includes three five-point Likert scale rating questions, one multiple choice question and one open-ended question.

\begin{table*}[h]
\centering
\caption{\small{Background of survey participants. Not Familiar (1), Slightly Familiar (2), Somewhat  Familiar (3), Knowledgeable (4), Expert (5)}}
\label{tab: bg}
\resizebox{\textwidth}{!}{
\small

\begin{tabular}{llll}
\toprule
\textbf{Role}                     & \textbf{Coding Experience (y)} & \textbf{Familiarity with ES-Products} & \textbf{Familiarity with \tooln} \\ \midrule
Researcher 65.6\% (21)            & 11--20 37.5\% (12)       & Somewhat familiar 40.6\% (13)         & Not familiar 40.6\% (13)                        \\
Software Developer 15.6\% (5)     & 6--10 28.1\% (9)         & Knowledgeable 28.1\% (9)              & Slightly familiar 25.0\% (8)\\
Student 12.5\% (4)                & 4--5 18.8\% (6)          & Slightly familiar 21.9\% (7)          & Somewhat familiar 18.8\% (6)                    \\
AI/ML Engineer 6.3\% (2)          & >21 12.5\% (4)            & Not familiar 6.3\% (2)                & Knowledgeable 15.6\% (5)                        \\
-                                 & 2--3 3.1\% (1)           & Expert 3.1\% (1)                      & Expert 0.0\% (0)\\ \midrule
\textbf{Familiarity with CodeGen}& \textbf{\# of Familiar CodeGen}& \textbf{Frequency of Using CodeGen}& \textbf{Experience of Working with CodeGen (y)}\\ \midrule
Knowledgeable 46.9\% (15)         & 4--6 50.0\% (16)& Daily 50.0\% (16)& 2 40.6.0\% (13)\\
Expert 28.1\% (9)                 & 2--3 25.0\% (8)& Weekly 28.1\% (9)                     & 3 31.3\% (10)                                   \\
Somewhat familiar 18.8\% (6)      & 7--10 15.6\% (5)         & Monthly 9.4\% (3)                     & 1 12.5\% (4)                                    \\
Slightly familiar 6.3\% (2)       & 11--20 6.3\% (2)         & Yearly 6.3\% (2)                      & 0 9.4\% (3)                                     \\
Not familiar 0.0\% (0)& 21+ 3.1\% (1)           & Never 6.3\% (2)                       & 4 6.3\% (2)                                     \\ \bottomrule
\end{tabular}

}
\end{table*}

\section{Survey Results}

We use Qualtrics to collect survey responses. 
For the open-ended questions, we use an open coding approach where two authors (coders) performed open coding on all the \numreplyvalid{} valid responses. They independently assigned one or more codes to each valid response. 
To assess inter-rater reliability for these multi-label annotations, we used the approach described in Section~\ref{sec: taxonomy}. The results show an overall label-level Cohen’s Kappa of 0.93 and an average Jaccard Index of 0.88, indicating a high level of inter-rater agreement.


\noindent\textbf{Background of Survey Participants.} 
Table \ref{tab: bg} shows the general demographic information of our survey participants. Most of our participants are researchers (65.6\%), have more than ten years of coding experience (50.0\%), are knowledgeable (46.9\%) about code generation models, and use the models frequently (daily (50.0\%) or weekly (28.1\%)). Most of them (75.0\%) are familiar with more than 3 code generation models, and 90.6\% of them have at least one year of experience working with code generation models. 
Most (40.6\%) are somewhat familiar with ethically sourced products, while the same proportion of participants are not familiar with \tooln. For the descriptions of different levels of familiarity, please refer to the specific question and options in our full survey.

\subsection{RQ1: Various Dimensions that Define \tooln}

Before asking participants to evaluate dimensions in our taxonomy, we collected the definitions provided by them via an open-ended question, and summarized the following dimensions:


\noindent\textbf{Overlapping Dimensions.} Most of the mentioned dimensions overlapped with our dimensions, indicating our collected taxonomy is general, and matches well with the stakeholders' perspective of \tooln. Notably, 46.9\% mentioned about licensing or copyright; 25.0\% mentioned environment; 21.9\% mentioned privacy (e.g., \emph{``Prompting data should not be stored, or should be stored securely, and deleted on demand''}), fairness (e.g., \emph{``minimizing bias''}), and source acknowledgment (e.g., some comments stated \emph{``to maintain transparency and traceability, we need to apply the concept of supply chain, like tracking the origin of code (i.e., which public code closely aligns with the generated code)''}, \emph{''The source for generating the code should be recorded and traced''}); and 18.8\% mentioned consent.



\noindent\textbf{Refined Aspect: Compensation to Authors of Input Training Data.} Three out of \numreplyvalid{} responses mentioned this so we merged it into ``fair wages'' under the ``labor rights'' dimension. As most code generation models mine data from the repository without gathering consent, this indicates \emph{the need to derive a scheme to provide compensation to authors of input training data}.
 

\noindent\textbf{Refined Aspect by Impacted Users: Opt-in Consent.} As impacted users spent effort to submit issues to opt-out from the Stack dataset, we notice that three users suggested about opt-in consent instead of the current opt-out consent policy. Notably, one comment defined \tooln{} as \emph{``Models should only be trained on code voluntarily offered up by developers, on an opt-in basis, and never scraped without consent, including even MIT licensed open source code, which was released as such before the proliferation of LLMs''}. 

    \noindent\textbf{New Dimension: Code Quality.} We did not include code quality as a dimension in our initial taxonomy because while metrics such as code quality and accuracy are important to assess the effectiveness of code generation models, they are typically regarded as technical quality indicators rather than ethical dimensions. However, 18.8\% of participants mentioned code quality (e.g., output quality, accuracy of generated code), with one mentioned \emph{``generation of seemingly correct but in fact faulty code, wasting developers' time in debugging''}(similar to ethical debates on ``hypocrite commit'' \cite{hypocrite2021} and nonsensical auto-generated commits \cite{linux_complain}). Another example directly related to ethical harm is generating source code with a bomb-making tutorial \cite{ren2024codeattack} where the auto-generated code can be used to encourage harmful behavior \cite{tan2025automated}. Thus, we added it in the taxonomy as a new dimension to offer a refreshed technical view of \tooln.


\begin{mybox}
\textbf{Finding 2:} Our summarized dimensions cover most of the dimensions in the definitions provided by the survey participants, with refined aspects consent (opt-in) and fair wage (compensation to authors of training data), and a new dimension on code quality (e.g., accuracy). We also observed that impacted users tend to mention opt-in consent when asked to define \tooln{}.
\end{mybox}

\begin{figure}[tbp]
\centering
\includegraphics[width=\linewidth]{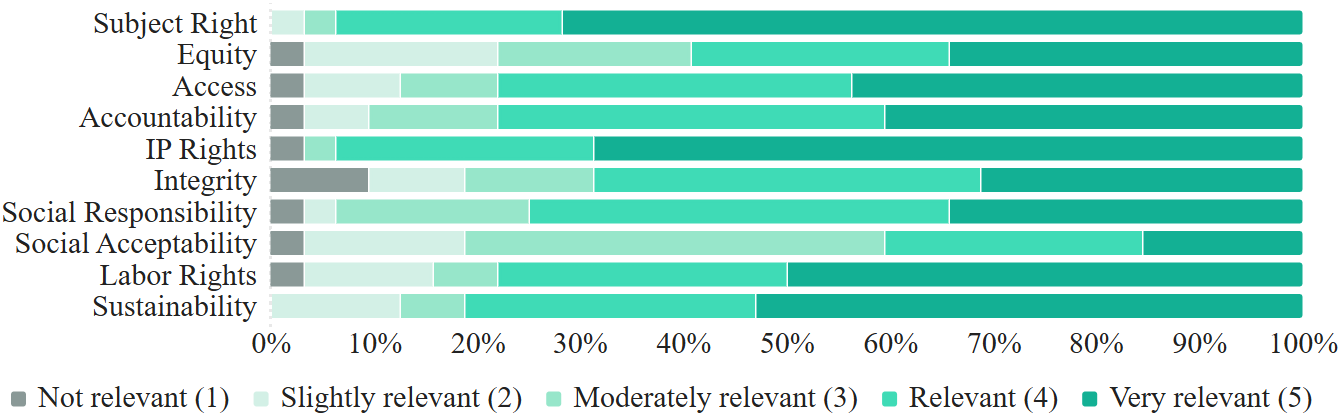}   
\caption{Relevance of Dimensions in Defining \tooln{}
}
\label{fig:evaluation_dimension}
\end{figure}

Figure \ref{fig:evaluation_dimension} shows the evaluation results on the dimensions from our taxonomy. The top-3 relevant dimensions participants considered are subject rights, intellectual property rights, and environmental sustainability. Overall, nearly all dimensions in our taxonomy are perceived as relevant or very relevant, with the three lowest-ranked dimensions, i.e., social acceptability (40.6\%) and equity (59.4\%) integrity (68.8\%), were still considered relevant or very relevant by many participants, suggesting that the dimensions in our taxonomy are all relevant to the definition of \tooln. 

\begin{mybox}
\textbf{Finding 3:} All dimensions in our taxonomy are considered relevant or very relevant, except for social acceptability. The most relevant ones are subject rights, intellectual property rights, and environmental sustainability.
\end{mybox}
When asked about the ignored dimensions before they took this survey, the top five dimensions are social acceptability (56.3\%), social responsibility (46.9\%), labor rights (46.9\%), access (28.1\%), and integrity (25.0\%). The least ignored dimension is intellectual property rights, which is consistent with their open-ended answers. Besides, 6.3\% of participants selected ``All of above'' while 12.5\% selected ``None of above'', indicating that the awareness of ethically sourced issues varies among stakeholders.

\begin{mybox}
\textbf{Finding 4:} Before taking the survey, participants tend to ignore dimensions from the social area when defining \tooln. However, most thought that these dimensions are at least moderately relevant in defining \tooln. 
\end{mybox}

\subsection{RQ2: Applicable Stages and Artifacts}


We have introduced the dimensions in Section~\ref{sec: taxonomy}. In RQ2, we asked participants to map the applicability of each dimension to various stages and artifacts involved in the supply chain of CodeGen (as shown in Figure \ref{fig:sc}). 

\underline{\noindent\emph{Stages.}}
We collected common stages in the development of CodeGen models from prior work \cite{mahmoudabadi2024promoting, schlegel2023management, wu2022sustainable}, and refined them to better examine where ethical risks may arise: 1) data collection
, 2) data annotation, 3) data cleaning and preprocessing, 4) model training and fine-tuning, 5) model evaluation, 6) deployment (integrate the trained model into practical environments, enabling it to perform tasks on designated systems either through real-time interaction or scheduled processing), 7) post-deployment (use model in real-world environment and monitoring its outputs and performance during inference). Specifically, we derive 7) post-deployment (Monitoring/Inference) by combining a) inference \cite{wu2022sustainable}, and b) model monitoring \cite{schlegel2023management}. 
Table~\ref{tab: stages} summarizes the applicability of each dimension to the various stages (based on the results of the survey). We can see that participants thought that all stages are generally applicable for all dimensions (on average, 39.4\% answered ``All of the above'' and at least one participant thought that each stage is applicable for each dimension). 
Among all stages, data collection is the most frequently selected to be applicable (50.0\% on average), with data annotation the second (40.3\%).

\begin{table*}[]
\centering
\caption{For each dimension, which stages of \tooln model development it can apply to?}
\small
\label{tab: stages}
\makebox[\textwidth]{
\resizebox{0.85\textwidth}{!}{
\begin{tabular}{lrrrrrrrrr}
\hline
Dimension                  & \multicolumn{1}{c}{\begin{tabular}[c]{@{}c@{}}Data \\ Collection\end{tabular}} & \multicolumn{1}{c}{\begin{tabular}[c]{@{}c@{}}Data \\ Annotation\end{tabular}} & \multicolumn{1}{c}{\begin{tabular}[c]{@{}c@{}}Data \\ Cleaning\end{tabular}} & \multicolumn{1}{c}{\begin{tabular}[c]{@{}c@{}}Model \\ Training\end{tabular}} & \multicolumn{1}{c}{\begin{tabular}[c]{@{}c@{}}Model \\ Evaluation\end{tabular}} & \multicolumn{1}{c}{\begin{tabular}[c]{@{}c@{}}Model \\ Deployment\end{tabular}} & \multicolumn{1}{c}{\begin{tabular}[c]{@{}c@{}}Post-\\ deployment\end{tabular}} & \multicolumn{1}{c}{\begin{tabular}[c]{@{}c@{}}None of \\ Above\end{tabular}} & \multicolumn{1}{c}{\begin{tabular}[c]{@{}c@{}}All of \\ Above\end{tabular}} \\ \hline
Subject Right (D1)         & 65.6\%                                                                         & 37.5\%                                                                         & 31.3\%                                                                       & 37.5\%                                                                        & 31.3\%                                                                          & 28.1\%                                                                          & 25.0\%                                                                         & 0.0\%                                                                        & 31.3\%                                                                      \\
Equity (D2)                & 50.0\%                                                                         & 34.4\%                                                                         & 34.4\%                                                                       & 34.4\%                                                                        & 43.8\%                                                                          & 31.3\%                                                                          & 25.0\%                                                                         & 0.0\%                                                                        & 37.5\%                                                                      \\
Access (D3)                & 40.6\%                                                                         & 31.3\%                                                                         & 28.1\%                                                                       & 28.1\%                                                                        & 37.5\%                                                                          & 46.9\%                                                                          & 37.5\%                                                                         & 0.0\%                                                                        & 34.4\%                                                                      \\
Accountability (D4)        & 46.9\%                                                                         & 43.8\%                                                                         & 40.6\%                                                                       & 40.6\%                                                                        & 50.0\%                                                                          & 43.8\%                                                                          & 31.3\%                                                                         & 0.0\%                                                                        & 40.6\%                                                                      \\
Intellectual Property (D5) & 65.6\%                                                                         & 46.9\%                                                                         & 34.4\%                                                                       & 40.6\%                                                                        & 34.4\%                                                                          & 31.3\%                                                                          & 25.0\%                                                                         & 0.0\%                                                                        & 34.4\%                                                                      \\
Integrity (D6)             & 56.3\%                                                                         & 37.5\%                                                                         & 46.9\%                                                                       & 34.4\%                                                                        & 43.8\%                                                                          & 21.9\%                                                                          & 25.0\%                                                                         & 3.1\%                                                                        & 31.3\%                                                                      \\
Social Responsibility (D8) & 50.0\%                                                                         & 40.6\%                                                                         & 37.5\%                                                                       & 34.4\%                                                                        & 37.5\%                                                                          & 28.1\%                                                                          & 28.1\%                                                                         & 3.1\%                                                                        & 40.6\%                                                                      \\
Social Acceptability (D9)  & 50.0\%                                                                         & 53.1\%                                                                         & 50.0\%                                                                       & 28.1\%                                                                        & 31.3\%                                                                          & 21.9\%                                                                          & 25.0\%                                                                         & 3.1\%                                                                        & 37.5\%                                                                      \\
Labor Right (D10)          & 40.6\%                                                                         & 50.0\%                                                                         & 37.5\%                                                                       & 25.0\%                                                                        & 25.0\%                                                                          & 21.9\%                                                                          & 25.0\%                                                                         & 6.3\%                                                                        & 50.0\%                                                                      \\
Sustainability (D11)       & 34.4\%                                                                         & 28.1\%                                                                         & 21.9\%                                                                       & 46.9\%                                                                        & 37.5\%                                                                          & 37.5\%                                                                          & 34.4\%                                                                         & 0.0\%                                                                        & 56.3\%                                                                      \\ \hline
Average                    & 50.0\%                                                                         & 40.3\%                                                                         & 36.3\%                                                                       & 35.0\%                                                                        & 37.2\%                                                                          & 31.3\%                                                                          & 28.1\%                                                                         & 1.6\%                                                                        & 39.4\%                                                                      \\ \hline
\end{tabular}
}}
\end{table*}


\begin{table*}[]
\centering
\caption{For each dimension, which artifacts of \tooln model development it can apply to?}
\small
\label{tab: artifacts}
\makebox[\textwidth]{
\resizebox{0.85\textwidth}{!}{
\begin{tabular}{lrrrrrrrr}
\hline
Dimension             & \multicolumn{1}{l}{Training Data} & \multicolumn{1}{l}{Dependencies} & \multicolumn{1}{l}{Model Metadata} & \multicolumn{1}{l}{Documentation} & \multicolumn{1}{l}{Prompt} & \multicolumn{1}{l}{Output} & \multicolumn{1}{l}{None of Above} & \multicolumn{1}{l}{All of Above} \\ \hline
Subject Right         & 68.8\%                            & 37.5\%                           & 28.1\%                             & 34.4\%                            & 15.6\%                     & 15.6\%                     & 0.0\%                             & 31.3\%                           \\
Equity                & 59.4\%                            & 15.6\%                           & 25.0\%                             & 37.5\%                            & 25.0\%                     & 53.1\%                     & 3.1\%                             & 21.9\%                           \\
Access                & 31.3\%                            & 34.4\%                           & 37.5\%                             & 53.1\%                            & 28.1\%                     & 37.5\%                     & 0.0\%                             & 40.6\%                           \\
Accountability        & 50.0\%                            & 46.9\%                           & 37.5\%                             & 50.0\%                            & 25.0\%                     & 31.3\%                     & 0.0\%                             & 43.8\%                           \\
Intellectual Property & 59.4\%                            & 40.6\%                           & 31.3\%                             & 34.4\%                            & 15.6\%                     & 15.6\%                     & 0.0\%                             & 37.5\%                           \\
Integrity             & 59.4\%                            & 31.3\%                           & 9.4\%                              & 21.9\%                            & 12.5\%                     & 21.9\%                     & 3.1\%                             & 31.3\%                           \\
Social Responsibility & 43.8\%                            & 31.3\%                           & 12.5\%                             & 28.1\%                            & 18.8\%                     & 28.1\%                     & 3.1\%                             & 43.8\%                           \\
Social Acceptability  & 50.0\%                            & 12.5\%                           & 18.8\%                             & 28.1\%                            & 21.9\%                     & 37.5\%                     & 0.0\%                             & 34.4\%                           \\
Labor Right           & 46.9\%                            & 28.1\%                           & 18.8\%                             & 21.9\%                            & 15.6\%                     & 18.8\%                     & 3.1\%                             & 46.9\%                           \\
Sustainability        & 40.6\%                            & 31.3\%                           & 28.1\%                             & 21.9\%                            & 18.8\%                     & 31.3\%                     & 3.1\%                             & 43.8\%                           \\ \hline
Average               & 50.9\%                            & 30.9\%                           & 24.7\%                             & 33.1\%                            & 19.7\%                     & 29.1\%                     & 1.6\%                             & 37.5\%                           \\ \hline
\end{tabular}
}}
\end{table*}

\underline{\noindent\emph{Artifacts.}} We also identified a set of artifacts---components within the model development and usage pipeline where ethically sourced issues may manifest from prior work \cite{schlegel2023management, aibomcomponent}: 
1) training data~\cite{garcia2025teaching,chowdhery2023palm}, 2) dependencies (i.e., software components including third-party libraries~\cite{aibomcomponent}), 3) model metadata (including learned parameters, hyperparameters~\cite{chowdhery2023palm}), 4) documentation~\cite{aibomcomponent}, 5) prompt (prompt design and template that may have bias or privacy risks~\cite{chowdhery2023palm,niu2023codexleaks}), 6) output (i.e., final product where ethical concerns become visible—ranging from biased code~\cite{feng2023investigating} to the inadvertent disclosure of private or confidential information, e.g., in the Samsung–ChatGPT incident \cite{samsung2023chatgpt}). Based on prior work \cite{schlegel2023management, aibomcomponent}, we additionally added ``prompt'' and ``output'' as new artifacts, as prior studies \cite{feng2023investigating, chowdhery2023palm, niu2023codexleaks} noted their importance.
Table~\ref{tab: artifacts} shows the applicability of each dimension to the various artifacts. Generally, the results are similar to those for the stages 
(each artifact has been noted as applicable by at least one participant). 
Among all artifacts, training data is the most frequently selected to be applicable  (50.9\% on average), with documentation the second (33.1\%). Meanwhile, 37.5\% thought that all artifacts are applicable, indicating that the entire supply chain of code generation model should be considered for \tooln{}.
\begin{mybox}
\textbf{Finding 5:} 
All stages and artifacts in the supply chain of \tooln are applicable to dimensions. This shows the importance of an ethical supply chain in \tooln.
\end{mybox}

\subsection{RQ3: Consequences of Un\tooln}
\subsubsection{Personal Experience}
Before and after rating our provided dimensions, we asked about participants' personal experience of facing ethical sourcing issues or situations where \tooln was needed. Before rating, 56.3\% said "no" for personal experience, and two impacted users noted consent, two noted licensing issues. After rating, only 37.5\% said ``no'', as most issues are covered by our dimensions. However, some practitioners adopt a more practical point of view of code generation where the usefulness of the code generation model is prioritized over other ethical dimensions, as two participants answered \emph{``As long as it functions well and passes the test I don't concern the sources''}. Besides, when asked about thoughts on dimensions, one pointed out potential trade-offs between dimensions, as \emph{``these dimensions often conflict in practice.  For example, you cannot achieve 100\% transparency and privacy simultaneously.''}


\subsubsection{Potential Impact}
 
We listed nine potential impacts collected from our literature review \cite{ma2024code, gupta2023chatgpt, majeed2024reliability, vartziotis2024learn, lau2023ban, stephanidis2025seven, yan2022whygen}, and asked participants to select all that apply. 
All listed impacts have at least one selection, and the detailed selection percentages are as follows: 

For model developers:
1. Lawsuit issues due to intellectual property violations (87.5\%), 
2. Security risks related to the misuse of personal data or lack of user consent  (87.5\%), 
3. Harmful outputs that reflect bias, toxic language, or offensive stereotypes  (78.1\%), 
4. Environmental impact (78.1\%), 
5. Reduced user trust and willingness to use the model (71.9\%), 
6. Negative public reputation (68.8\%), 
7. Exclusion of certain users or communities due to limited access or high costs (46.9\%). 
For users:
8. Lawsuits due to using copyrighted code generated by models (90.6\%), 
9. Criticized for plagiarism due to using generated code that is too similar to existing code (75.0\%). 
This result is consistent with Finding 3 --- most participants cared about intellectual property rights and subject rights.

\noindent\textbf{Open-ended Questions.} Before presenting the options, we asked an open-ended question about potential impacts and collected the following new impacts: 1. Exploitation of open-source developers (15.6\%, as one comment stated \emph{``human developers not getting the recognition or compensation they deserve''}), 2. Low-quality or unreliable generated code (12.5\%, e.g., \emph{``buggy code''}), 3. \emph{``Monopolization of generative AI''} (9.4\%), 4. Impact on open-source communities (9.4\%, as \emph{``more projects being closed source to avoid scraping'', ``The tragedy of the commons. Less people will contribute to the commons''}) 5. Reliance on CodeGen models (6.3\%), as one noted in `any comment': \emph{``User-cognitive (effects of reliance on code generation models, dependency)''}, 6. Fuel of dishonesty (3.1\%), 7. Unemployment (3.1\%). 


\begin{mybox}
\textbf{Finding 6:} Among the potential impact options, most participants cared about the lawsuit issues due to IP violations on users (90.6\%) and model developers (87.5\%), then security risks on model developers (87.5\%). The top-3 new important impacts noted include 1. Exploitation of developers (15.6\%), 2. Low-quality generated code (12.5\%), 3. Monopolization of generative AI and impact on open-source communities (9.4\%). 
\end{mybox}

\subsection{RQ4: Trade-offs \& Contamination}






\noindent\textbf{Ranking Trade-offs.} As noted in studies from our literature review~\cite{chowdhery2023palm, barocas2021designing}, developing CodeGen models needs to balance key trade-offs.  Hence, we asked participants to rank the relative importance of five potential trade-offs for \tooln, with rank=1 the most important. The result shows that the top-ranked is ``accuracy loss'' (average rank 1.84), followed by ``response time or latency'' (2.69), ``data contamination'' (3.09), ``financial or resource cost'' (3.28), and ``data preparation time'' (4.09). 

\begin{figure}[tbp]
\centering
\includegraphics[width=\linewidth]{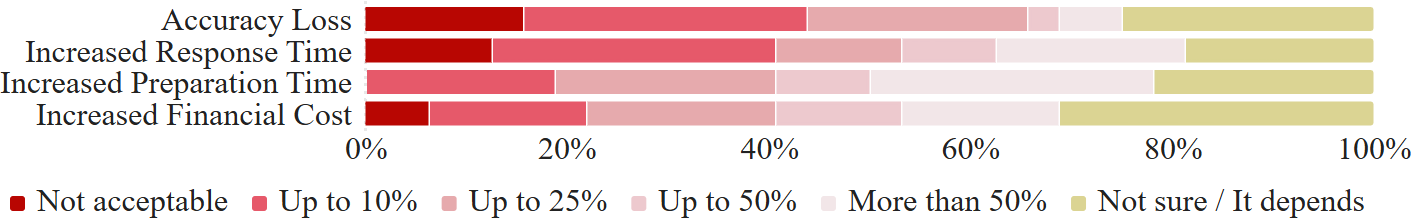}   
\caption{Extent of Acceptable Trade-offs
}
\label{fig:cost}
\end{figure}

\noindent\textbf{Acceptable Extents.} Figure \ref{fig:cost} shows the extent to which participants are willing to accept each trade-off (except for contamination), in terms of percentage loss or increase. For accuracy loss (top-ranked previously), 15.6\% selected that no loss is acceptable, and  28.1\% selected that up to 10\% loss is acceptable. For data preparation time (lowest-ranked previously), 28.1\% selected that more than 50\%, 9.4\% selected up to 50\%, 
and no one selected unacceptable. 

\noindent\textbf{Acceptable Contamination.} We provided six contamination types derived from artifacts and let them select all that apply. As a result, the most acceptable ones are automatically generated content (53.1\%) and bot submissions (34.4\%), followed by data with unclear or unknown dependencies (28.1\%), using closed-source models (28.1\%), unknown or unrevealed data sources (25.0\%), and data with incompatible license within the dependencies (25.0\%). Among all, 31.3\% of participants thought no contamination is acceptable.


\begin{mybox}
\textbf{Finding 7:} The most unacceptable trade-off is accuracy loss (most participants (28.1\%) thought that acceptable accuracy loss is up to 10\%), while the least is data preparation time. The most unacceptable contamination types are unrevealed data sources and data with incompatible licenses within the dependencies. 31.3\% thought that no contamination is acceptable.
\end{mybox}

\subsection{Ethically Source Code Generation Models}
After learning about the various dimensions, most participants (68.8\%) thought that none of the existing code generation models aligns with the definition of \tooln. Meanwhile, 21.9\% felt that some tools partially aligned, while two participants answered "yes" and one indicated ``no idea''. Some practitioners provided examples of partially ethically sourced models, including Llama, Qwen, and Anthropic Claude Code, but these models still need to improve on dimensions like transparency and opt-in consent.

Those who answered ``yes'' mentioned StarCoder and OLMo2, as one impacted user said \emph{``OLMo2 uses ethically sourced data and doesn't use data without consent while still getting very good performance. They also list all the licenses of which they are using the data under''}. Based on the OLMo2's report~\cite{olmo20242}, we notice that as OLMo2 also relied on the Stack dataset, it may share the similar problem of opt-in consent. Among the six impacted users, three answered ``no'', two noted ``partially'' (e.g., \emph{``Most tools still lack full transparency and licensing clarity. However, tools that provide citation, allow opt-outs for training data, or offer explainability features are moving in the right direction...'')}, and one noted ``yes'', showing that most impacted users are unsatisfied with the practices of current models. 

\begin{mybox}
\textbf{Finding 8:} Most participants thought that none of the currently available CodeGen models aligns with the definition of \tooln or only partially aligns, suggesting a considerable gap between current models and \tooln.
\end{mybox}

\subsection{Post-Survey Reflections}
More than half of the participants agreed (56.3\%) or strongly agreed (18.8\%) that their understanding of \tooln has improved, with the remaining (25.0\%) selecting neutral.
Similarly, most participants agreed (43.8\%) or strongly agreed (15.6\%) that this survey made them feel that the issues of \tooln were more important than they previously thought. 37.5\% selected neutral, and only 3.1\% disagreed.
Finally, most participants believed that the dimensions and definitions discussed in this survey could be generalized to other fields beyond code generation (e.g., image generation, voice generation). Specifically, 46.9\% chose probably yes (4), 43.8\% chose definitely yes (5), and 9.4\% chose not sure. 

\begin{mybox}
\textbf{Finding 9:} Our survey helped improve more than half of the participants' understanding of \tooln, raised awareness, and affirmed the generalizability of the definition of \tooln.
\end{mybox}
\section{Discussions and Implications}
We highlight several implications for the research community:

\noindent\textbf{Multidimensional View in Defining \tooln{}.} Our literature review identified 10 dimensions that define \tooln{} (Finding 1), whereas our survey further refined the ``labor rights'' dimension by stressing the compensation to authors of training data (Finding 2). Moreover, participants acknowledged that they tend to ignore dimensions in the social area (social acceptability, social responsibility, and labor rights) (Finding 4). As all these dimensions are considered relevant or very relevant in defining \tooln (Finding 3), our study indicates \emph{the need to adopt a multidimensional view in defining \tooln}, and \emph{help raise awareness about various dimensions towards \tooln} (Finding 9).

\noindent\textbf{Code Quality as a New Technical Dimension of \tooln.} Stakeholders have expressed the importance of incorporating code quality as a dimension of \tooln (Finding 2), and indicated the impact of low quality code being generated (Finding 6). We believe that this \emph{offers a refreshed technical view of \tooln{}, and stresses the importance of adopting technical lens} while considering ethical dimensions of \tooln. 

\noindent\textbf{Importance of Opt-in Consent Noted by Impacted Users.} We observe that impacted users who are directly affected by Un\tooln are more willing to express their opinions on \tooln. Instead of taking initiative to opt-out from the Stack dataset, they also preferred opt-in consent (Finding 2). This shows that \emph{simply enforcing informed consent may be insufficient} as a specific mechanism (opt-in) is preferred over existing practices (opt-out). As data collection for training Code LLMs usually involves mining data from millions of repositories, and these repository owners may not provide their contact information, it is challenging to obtain opt-in consent from these repository owners. Hence, for open-source repository owners, they should consider adding Non-AI licenses to their repositories \cite{nonAILicenses} to explicitly express disagreement with the use of their data for AI training. For AI developers or Data miners, they should  
\emph{design practical yet ethical approaches to obtain large-scale opt-in consent from developers}, such as developing a consent-management system for open-source repositories.

\noindent\textbf{Consequences of Un\tooln.} Stakeholders noted that there are many potential consequences of Un\tooln where most participants cared about lawsuit issues that may be faced by users and model developers (Finding 6). Participants also suggested new potential impacts from the perspective of developers and users: 1. Exploitation of developers, 2. Low-quality generated code, 3. Monopolization of generative AI, and 4. Impact on open-source communities. These consequences \emph{help strengthen the motivations of our study} and \emph{raise awareness of the importance of \tooln}.

\noindent\textbf{Ethically Sourced Supply Chain in \tooln.} 
As Finding 5 revealed that 
all stages and artifacts in the supply chain of \tooln are considered to be relevant to the dimensions,
our study \emph{serves as motivation to ensure that the entire supply chain of code generation is ethically sourced}. Moreover, we also observe that most studies focus on evaluating the generated output \cite{yan2022whygen,feng2023investigating, liu2024toward,ling2024evaluating, liu2023uncovering}, neglecting other important stages (e.g., data annotation) and artifacts (e.g., documentation). In the future, it is worthwhile to systematically investigate the ethical perspectives of these important stages and artifacts.

\noindent\textbf{Acceptable Trade-offs and Contamination.} As most participants noted that code quality is an important dimension, Finding 7 revealed that most noted that accuracy loss is unacceptable (they can only accept up to 10\% loss). Some  thought that no contamination is acceptable, whereas most thought that they cannot accept contamination from unknown or unrevealed data sources and data with incompatible licenses within the dependencies. As removing data contamination may result in less data that may in turn lead to accuracy loss, there is \emph{a need to balance the trade-off between accuracy and other dimensions (e.g., informed consent and IP rights).} 

\noindent\textbf{Gaps in CodeGen Models.} Finding 8 revealed that most participants thought that none of the current CodeGen models aligns with the definition of
\tooln or only partially aligns. Practitioners noted that current models still need to improve on dimensions like transparency and opt-in consent, indicating the \emph{gaps in existing code generation models}. This calls for the need to design practical techniques to improve CodeGen models on these dimensions. 
\section{Threats to Validity}
\label{threat}

\noindent\textbf{External.} Our study derives the taxonomy based on our literature review and survey. As ethical issues are trending topics, additional newer literature might have exposed more dimensions in defining \tooln. As most dimensions in our taxonomy overlaps with the participants' definitions, our survey results show that our taxonomy are quite general and comprehensive. Our findings may not generalize beyond the survey results provided by the \numreplyvalid responses. We mitigate this by sending multiple rounds of invitations to obtain diverse options from various stakeholders.  

\noindent\textbf{Internal.} 
 To reduce researcher bias in open-ended
 response coding, we use an iterative coding process for all disagreements to reach a consensus. To ensure that we covered stakeholders with
 diverse backgrounds, we used diverse strategies to
 invite participants (e.g., posting to mailing lists, contacting stakeholders through our professional network, and emailing impacted users by searching for submitted GitHub issues). However, the low response rate and self-selection bias may have affected the results by attracting participants interested in the survey topic. We designed
 our survey questions by following best practices and survey design guidelines. 
\section{Conclusion}

This paper introduces the notion of \tooln{} that encompasses 11 different ethical dimensions. To build a broad yet relevant taxonomy of \tooln{}, we perform a two-phase literature review where we reviewed \manualread{} papers to identify \relevantpaper{} relevant papers with 10 initial dimensions of \tooln. Based on our literature review, we surveyed \numreplyvalid{} practitioners to identified various dimensions, consequences, artifacts, and stages relevant to \tooln{}, leading to 11 dimensions of \tooln{} with new dimension focusing on quality. The evaluation of our survey showed most practitioners either agreed or strongly agreed that our survey help improve their understanding of \tooln{}. As practitioners noted that none of existing code generation models aligns with the definition of \tooln,
our findings call for the need to design approaches towards \tooln{}.

\section{Acknowledgement}
We acknowledge the support of the Government of Canada’s New Frontiers in Research Fund (NFRF), [NFRFE-2024-00612], the insightful suggestions from reveiwers, and detailed answers from all survey participants. We also express our sincere gratitude to Professor Yu Huang and Professor Diego Elias Costa for their invaluable help in this research. 


\balance
\bibliographystyle{ACM-Reference-Format}
\bibliography{ref}


\end{document}